\documentclass{article}
\usepackage[T1]{fontenc} 
\usepackage[utf8]{inputenc} 
\usepackage{ismir,amsmath,cite,url}
\usepackage{graphicx}
\usepackage{color}

\title{Towards Automatic Instrumentation by Learning to Separate Parts in Symbolic Multitrack Music}

\multauthor
{Hao-Wen Dong$^1$ \hspace{1em} Chris Donahue$^2$ \hspace{1em} Taylor Berg-Kirkpatrick$^1$ \hspace{1em} Julian McAuley$^1$}
{$^1$ University of California San Diego \hspace{1em} $^2$ Stanford University\\\small$^1$\texttt{\{\href{mailto:hwdong@ucsd.edu}{hwdong},\href{mailto:tberg@ucsd.edu}{tberg},\href{mailto:jmcauley@ucsd.edu}{jmcauley}\}@ucsd.edu} \hspace{1em} $^2$ \texttt{\href{mailto:cdonahue@cs.stanford.edu}{cdonahue@cs.stanford.edu}}
}

\def\authorname{Hao-Wen Dong, Chris Donahue, Taylor Berg-Kirkpatrick and Julian McAuley}

\usepackage[bookmarks=false,pdfauthor={\authorname},pdfsubject={\papersubject},hidelinks]{hyperref}

\usepackage{booktabs}
\usepackage{amsfonts}
\usepackage[capitalise,noabbrev]{cleveref}
\usepackage{enumitem}
\usepackage{bbm}
\usepackage{multirow}
\usepackage{tabularx}
\usepackage{xcolor}
\usepackage{refcount}
\usepackage{arydshln}
\usepackage{overpic}

\graphicspath{{figs/}}

\DeclareMathOperator*{\argmin}{arg\,min}
\setlist[itemize]{leftmargin=*,itemsep=.5ex,topsep=0pt,partopsep=0pt,parsep=0pt}

\crefformat{footnote}{#2\footnotemark[#1]#3}

\definecolor{tab-blue}{HTML}{1f77b4}
\definecolor{tab-orange}{HTML}{ff7f0e}
\definecolor{tab-green}{HTML}{2ca02c}
\definecolor{tab-red}{HTML}{d62728}
\definecolor{tab-purple}{HTML}{9467bd}

\begin{document}

\maketitle

\begin{abstract}
Modern keyboards allow a musician to play multiple instruments at the same time by assigning zones---fixed pitch ranges of the keyboard---to different instruments. In this paper, we aim to further extend this idea and examine the feasibility of automatic instrumentation---dynamically assigning instruments to notes in solo music during performance. In addition to the online, real-time-capable setting for performative use cases, automatic instrumentation can also find applications in assistive composing tools in an offline setting. Due to the lack of paired data of original solo music and their full arrangements, we approach automatic instrumentation by learning to separate parts (e.g.,~voices, instruments and tracks) from their mixture in symbolic multitrack music, assuming that the mixture is to be played on a keyboard. We frame the task of part separation as a sequential multi-class classification problem and adopt machine learning to map sequences of notes into sequences of part labels. To examine the effectiveness of our proposed models, we conduct a comprehensive empirical evaluation over four diverse datasets of different genres and ensembles---Bach chorales, string quartets, game music and pop music. Our experiments show that the proposed models outperform various baselines. We also demonstrate the potential for our proposed models to produce alternative convincing instrumentations for an existing arrangement by separating its mixture into parts. All source code and audio samples can be found at \url{https://salu133445.github.io/arranger/}.
\end{abstract}

\section{Introduction}

\begin{figure}
    \centering
    \includegraphics[width=\linewidth]{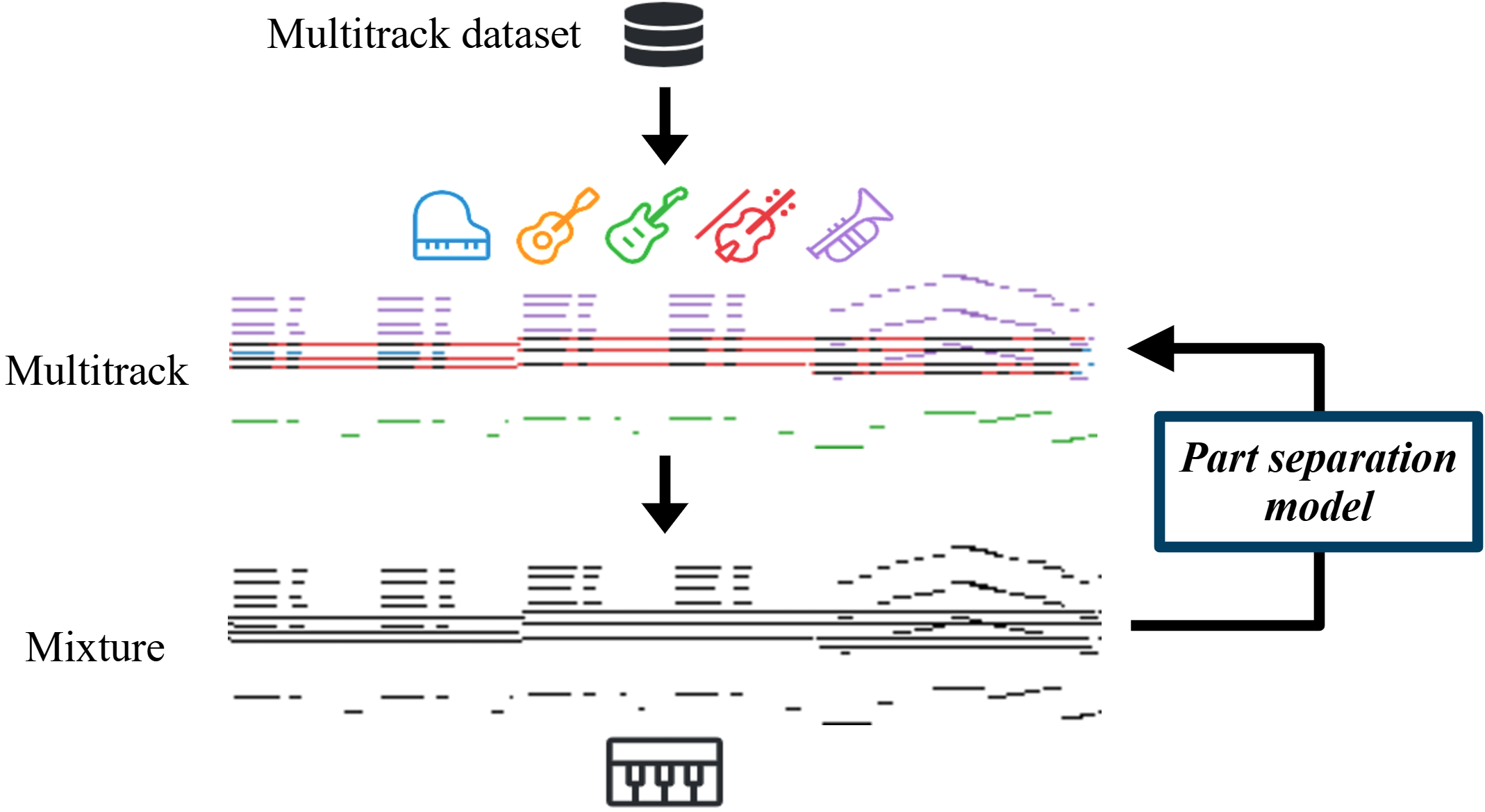}\\[-3ex]
    \raggedleft{\small(Audio available.\cref{fn:demo})}
    \caption{Proposed pipeline. By downmixing a symbolic multitrack into a single-track mixture, we acquire paired data of solo music and its instrumentation. We then use these paired data to train a \textit{part separation} model that aims to infer the part label (e.g., one out of the five instruments in this example) for each single note in a mixture. Automatic instrumentation can subsequently be accomplished by treating input from a keyboard player as a downmixed mixture (bottom) and separating out the relevant parts (top). The music is visualized in the piano roll representation, where the x- and y-axes represent time and pitch, respectively. Colors indicate the instruments.}
    \label{fig:pipeline}
\end{figure}

\begin{figure*}
    \centering
    \small
    \def\imgheight{8.5ex}
    \begin{tabularx}{\linewidth}{m{.075\linewidth} m{.875\linewidth}}
        Musical score &\includegraphics[width=\linewidth]{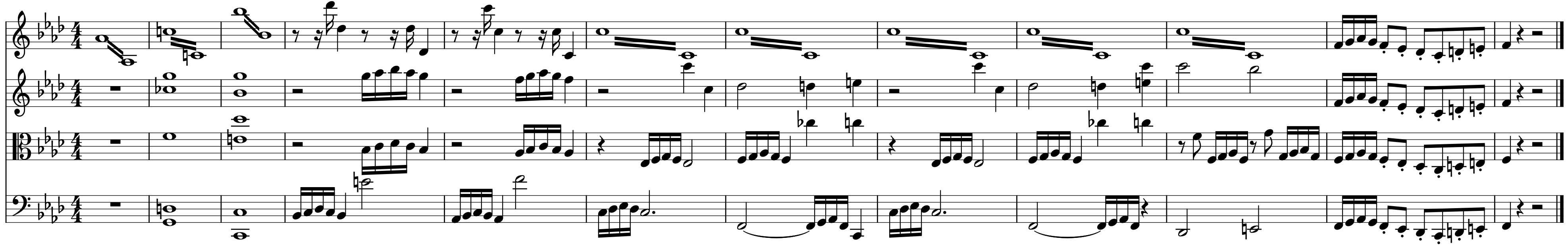}\\
        \midrule
        Mixture (input) &\includegraphics[width=\linewidth,height=\imgheight]{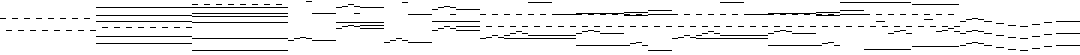}\\
        \midrule
        Ground truth &\includegraphics[width=\linewidth,height=\imgheight]{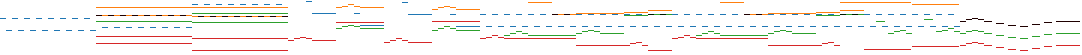}\\
        \midrule
        Online LSTM\hspace{1ex} prediction &\includegraphics[width=\linewidth,height=\imgheight]{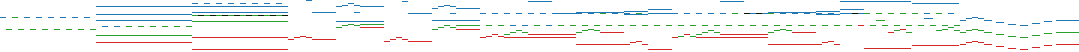}\\
        \midrule
        Offline BiLSTM prediction &\includegraphics[width=\linewidth,height=\imgheight]{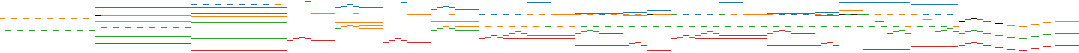}\\
        \multicolumn{2}{r}{\footnotesize (Audio available.\cref{fn:demo} Colors: \textcolor{tab-blue}{first violin}, \textcolor{tab-orange}{second violin}, \textcolor{tab-green}{viola}, \textcolor{tab-red}{cello}.)}
    \end{tabularx}
    \caption{Hard excerpt in the string quartets dataset---Beethoven's \textit{String Quartet No. 11 in F minor, Op. 95,} movement 1, measures 72--83. The tremolos of the first violin (measures 1--3 and 6--10), the double stops for the second violin, viola and cello (measures 2--3) and the overlapping pitch ranges (measures 2--5) together compose a complex texture. Both models fail to handle the violins and viola properly, especially the second violin.}
    \label{fig:sample_musicnet}
\end{figure*}

Music is an art of time and sound. It often contains complex textures and possibly \emph{parts} for multiple voices, instruments and tracks. While jointly following the global style and flow of the song, each part possesses its own characteristics and can develop different musical ideas independently. For example, in pop music, guitar and piano tend to play chords and might span across a large pitch range, while bass is usually monophonic and stays in a lower range. While playing multiple instruments usually requires multiple performers, keyboardists potentially have the ability to control many instruments at once. Modern keyboards often offer the functionality of \emph{zoning}, which allows a player to divide the pitch range into zones and assign each zone to a certain instrument. However, zoning is not ideal given its low flexibility that requires careful configuration and sometimes rearrangement of the music, and incapability for handling certain genres of music that have close and possibly overlapping harmony.

In this paper, we aim for the more ambitious goal of automatic instrumentation---a process that we define as dynamically assigning instruments to notes in solo music. A real-time, online automatic instrumentation model could allow a musician to have their keyboard performance instantaneously and seamlessly performed by a different ensemble. In addition to performative use cases, an offline automatic instrumentation model can also be useful to assist composers in suggesting proper instrumentation or providing a starting point for arranging a solo piece, especially for composers who have less experience arranging for a particular ensemble.

Automatic instrumentation is challenging as it requires domain knowledge of each target instrument, i.e., which pitches, rhythms, chords, and sequences thereof are playable, and it is hard to specify such knowledge by some fixed set of rules. In view of recent advances in machine learning, we propose to adopt a data-driven approach to this task. However, it can be laborious to acquire paired data of original solo music and their full arrangements. Given the abundance of multitrack music data, we approach automatic instrumentation by learning to separate parts from their mixture in multitrack music, a task we call \emph{part separation} (see \cref{fig:pipeline}). Assuming that the mixture is to be played on a keyboard and the multitrack is the target arrangement that we want to generate by an automatic instrumentation model, we thereby have paired data for solo music and full arrangements.

We frame the new task of part separation as a sequential multi-class classification problem that aims to map sequences of notes into sequences of part labels. We adopt long short-term memory~(LSTM)~\cite{hochreiter1997lstm} and Transformer~\cite{hsiao2021transformer} models for the task. We conduct an extensive empirical evaluation showing the superiority of our proposed models to baselines for the related task of voice separation as well as strategies found in commodity keyboards. To showcase the potential of our proposed models, we also demonstrate their ability to produce alternative convincing instrumentations for existing arrangements. Audio for all examples and more samples are available on the demo website.\footnote{\url{https://salu133445.github.io/arranger/}\label{fn:demo}} All source code can be found in the project repository.\footnote{\url{https://github.com/salu133445/arranger}}

\section{Prior Work}

Voice separation is a related task to part separation which involves separating blended scores into individual monophonic voices. While useful, voice separation is agnostic to constraints imposed by specific instruments---a composer using a voice separation algorithm would have to manually align voices to appropriate instruments. Some prior work investigates voice separation in small, carefully-annotated pop music datasets~\cite{gray2016voiceseparation,guiomardkagan2015voice}. Some prior work on voice separation allows synchronous or overlapping notes in a voice~\cite{kilian2002voiceseparation,cambouropoulos2006voice,karydis2007visa,cambouropoulos2008voice}. However, their results are only reported on small test sets in certain genres. Others have adopted multilayer perceptrons~\cite{gray2016voiceseparation,devalk2019deepseparation} and convolutional neural networks~\cite{gray2020voiceseparation} with hand-crafted input features for voice separation. Another relevant work on hand detection in piano music used LSTMs to separate notes played by right and left hands in piano MIDI data~\cite{hadjakos2019hands}. To the best of our knowledge, no past work has examined the task of part separation in a general setting for multiple music genres.

In addition to voice separation, prior work has explored automatic music arrangement. 
The primary focus of prior work for automatic music arrangement has been on reduction---mapping musical scores for large ensembles to parts playable by a single specific instrument such as the piano~\cite{chiu2009automatic,onuma2010piano,huang2012towards,nakamura2015automatic,takamori2017automatic,nakamura2018statistical}, guitar~\cite{tuohy2005genetic,hori2012automatic,hori2013input} or bass~\cite{abe2012automatic}. This past work focuses on identifying the least important notes to delete so that the resultant score is playable on a single instrument, whereas our work seeks to preserve the original score in its entirety and satisfy playability for multiple instruments simultaneously. As an exception, Crestel and Esling~\cite{crestel2016live} explore strategies for arranging orchestral music from piano, though their approach does not guarantee that all notes in the input piano map to parts in the output.

Music generation is another body of work that has used neural network sequential models for processing symbolic music~\cite{briot2017survey}. 
Simon and Oore~\cite{simon2017performance} proposed a convenient approach for music generation which involved training recurrent neural network language models on a language-like ``event-based'' representation of music. 
Subsequently, recent work has explored event-based representations using Transformers~\cite{huang2019musictransformer,musenet,donahue2019lakhnes,huang2020pop,ens2020mmm,hsiao2021transformer,muhamed2021transformergan}. 
In this work, we explore a more compact input representation of music that passes all of the information about a note into the model at once, rather than spreading it out across several events.
We also note that Payne~\cite{musenet} generate music which contains parts for several instruments, but their model cannot be directly used to perform part separation of existing musical material.

\section{Problem Formulation}

Mathematically, we consider a piece of music $x$ as a sequence of notes $(x_1,\dots,x_N)$, where $N$ is the number of notes. Each note is represented by a tuple of time $t_i$ and pitch $p_i$, i.e., $x_i = (t_i, p_i)$. Alternatively, we could also include duration $d_i$ as an input and have $x_i = (t_i, p_i, d_i)$. Each note is associated with a label $y_i \in \{1,\dots,K\}$ that represents the part it is in, where $K$ is the number of parts. The goal of part separation is to learn the mapping between notes and part labels. This formulation is rather flexible and has no assumptions on whether a part is monophonic or not---it can be a voice, an instrument, a track, etc.

In terms of the context given for predicting the label of each note, we can categorize part separation models into three classes: An \emph{independent model} predicts the label for each note independently, without any context. An \emph{online model} predicts the label of the current note $x_i$ given only past information, i.e., notes $(x_1,\dots,x_{i-1})$, as context. An \emph{offline model} predicts the label of the current note $x_i$ given past and future information, i.e.,~the full sequence of $(x_1,\dots,x_N)$, as context. While independent and online models are preferable for use cases that require real-time outputs,~e.g., live performance. Moreover, the inability to look into the future makes the real-time setting more challenging than the offline setting. On the other hand, offline models can find applications in assisstive composing tools.

\begin{table*}
    \centering
    \small
    \begin{tabular}{llllllll}
        \toprule
        Dataset &Hours &Files &Notes &Parts &Ensemble &Most common label\\
        \midrule
        Bach chorales~\cite{cuthbert2010music21}     &3.23  &409   &96.6K &4 &soprano, alto, tenor, bass                 &bass (27.05\%)\\
        String quartets~\cite{thickstun2017musicnet} &6.31  &57    &226K  &4 &first violin, second violin, viola, cello  &first violin (38.72\%)\\
        Game music~\cite{donahue2018nesmdb}          &45.05 &4.61K &2.46M &3 &pulse wave I, pulse wave II, triangle wave &pulse wave II (39.35\%)\\
        Pop music~\cite{raffel16thesis}              &1.02K &16.2K &63.6M &5 &piano, guitar, bass, strings, brass        &guitar (42.50\%)\\
        \bottomrule
    \end{tabular}
    \caption{Statistics of the four datasets considered in this paper.}
    \label{tab:data}
\end{table*}

\section{Models}
\label{sec:model}

We consider the following input features for our models---(1) \emph{time}: onset time, in time step,\footnote{Assuming that the music is in metrical timing, a time step is a factor of some musically-meaningful unit (e.g., a quarter note) and can be adjusted to match the desired temporal resolution.} (2) \emph{pitch}: pitch as a MIDI note number, (3) \emph{duration}: note length, in time step, and (4) \emph{frequency}: fundamental frequency of the pitch, in Hz, computed by the formula $f = 440 \cdot 2^{(p - 69)/12}$. In addition, we also consider features that encode the metric time grid of music similar to the \textsc{Bar} and \textsc{Position} events proposed in~\cite{huang2020pop}---(5) \emph{beat}: onset time, in beat, and (6) \emph{position}: position within a beat, in time step.

Moreover, to help the models better disambiguate parts, we also include two simple hints---(7) \emph{entry hints}: onset position for each instrument, encoded as a unit step function centered at its onset time and all zero if the instrument is not used, and (8) \emph{pitch hints}: average pitch of each track. These hints allow the musician to use interactively to make the instrumentation process more controllable. For example, entry hints can be used to control the instruments available as they serve as switches for the instruments.

For the machine learning models, we consider the LSTM~\cite{hochreiter1997lstm} and its bidirectional version (BiLSTM)~\cite{schuster1997bilstm}. We use a three-layer stacked LSTM with 128 hidden units in each layer (64 hidden units per layer for BiLSTM). We also consider two variants of Transformer~\cite{vaswani2017transformer}---one based on the encoder (Transformer-Enc) and one based on the decoder (Transformer-Dec). They share the same architecture that is composed of three Transformer blocks, each of which has 128 hidden units and 8 heads in self-attention computation and 256 hidden units in the internal feedforward network. However, they have different attention masks: Transformer-Enc uses only the padding mask, while Transformer-Dec uses both the padding mask and the lookahead mask, which blocks its access to future information and makes it a online model. In this paper, the LSTM and Transformer-Dec models are made online models, and the BiLSTM and Transformer-Enc models are made offline models that take durations as inputs.

\section{Baseline Models}

In order to gain an insight into how the proposed models perform, we include two heuristic algorithms and a voice separation model from the literature in our empirical study.

\subsection{Zone-based algorithm}

This algorithm simulates a common feature in modern keyboards where a player can preassign a pitch range (i.e., the `zone') for each instrument and notes will automatically be assigned to the corresponding instrument as the player performs. This algorithm finds the optimal zones for the whole training data and uses these optimal zones at test time. For the oracle case, the optimal zones for each sample are computed and used at test time. We note that the oracle case might not be easily achievable as it can be hard for a musician to set the zones optimally beforehand, especially for improvisation.

\subsection{Closest-pitch algorithm}

The closest-pitch algorithm keeps track of the last active pitches $p'_i$ for each track $i$. For each incoming pitch $p$, it finds the pitch among the last active pitches that has the closest pitch to $p$ and assigns the upcoming note with the same label as the chosen pitch. This is a casual model and it also relies on the onset hints. We can formulate this algorithm as follows. For $i = 1,\dots,N$, we have $$\hat{y}_i = \begin{cases}y_i, &\text{if $x_i$ is an onset}\\\displaystyle\argmin_{j\in\{1,\dots,K\}} (p_i - p'_j)^2 + M a_i, &\text{otherwise}\end{cases}\,,$$ where $p'_i$ is the last active pitch of track $i$ before time $t$ and $a_i$ indicates whether track $i$ is active, i.e., a concurrent note has not yet been released. We set $M$ to a large positive number when we assume each part is monophonic, which we will refer to as the `mono' version of this algorithm, otherwise set $M=0$.

\subsection{Multilayer perceptron (MLP)}

We adapt the voice separation model proposed in~\cite{devalk2019deepseparation} to the task of part separation. This model uses multilayer perceptron (MLP) to predict the label for the current note based on hand-crafted features that encodes its nearby context. We use entry hints rather than predicting them by the proposed voice entry estimation heuristics. We remove the `interval' feature as there is no upper bound for the number of concurrent notes and change the proximity function to L1 distance. The oracle case of this model replaces error-prone prior predictions with ground truth history labels. In our implementation, we use three fully-connected layers with 128 hidden units each.

\section{Data}

In order to examine the effectiveness of the proposed models, we consider four datasets---(1) \emph{Bach chorales} in Music21~\cite{cuthbert2010music21}, (2) \emph{string quartets} in MusicNet~\cite{thickstun2017musicnet}, (3) \emph{game music} in Nintendo Entertainment System (NES) Music Database~\cite{donahue2018nesmdb} and (4) \emph{pop music} in Lakh MIDI Dataset~\cite{raffel16thesis}, which are diverse in their genres, sizes and ensembles (see~\cref{tab:data} for a comparison).

As these datasets are noisy in different ways, we need to further clean the data. For the game music dataset, we discard the percussive noise track in the original dataset as they do not follow the standard 128-pitch system used in other tracks. For the pop music dataset, we use a cleaned subset derived in~\cite{dong2018musegan}, which contains only pop songs. We mapped the instruments to the five most common instrument families---piano, guitar, bass, strings and brass. We follow the General MIDI 1 specification on the mapping from an instrument to its instrument family. Instruments that fall outside of these five families are discarded. We note that the lead melody track might occasionally be discarded during the mapping process due to the high variance on instruments used for the melody track.

Moreover, we discard songs with only one active track as the task becomes trivial in this case. We note that all Bach chorales, string quartets and most pop songs are in metrical timing, where a time step corresponds to some fraction of a quarter note. Thus, we downsample them into 24 time steps per quarter note, which can cover 32nd notes and triplets. As songs in the game music dataset are in absolute time, we downsample them to a temporal resolution equivalent to 24 time steps per quarter note in a tempo of 125 quarter notes per minute (qpm).

Finally, we split each dataset into train--test--validation sets with a ratio of $8 : 1 : 1$ except the game music dataset, where we use the original splits provided with the NES Music Database. We use MusPy~\cite{dong2020muspy} and music21~\cite{cuthbert2010music21} for processing MIDI and MusicXML files.

\begin{figure}
    \centering
    \small
    \def\imgheight{5ex}
    \begin{tabularx}{\linewidth}{m{.16\linewidth} X}
        Musical score
          &\includegraphics[width=\linewidth]{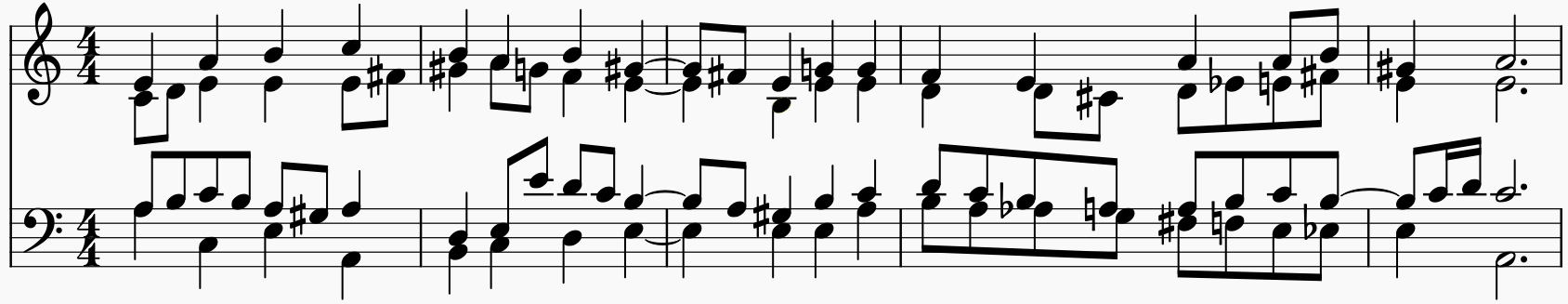}\\
        \midrule
        Ground truth
          &\includegraphics[width=\linewidth,height=\imgheight]{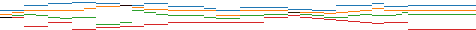}\\
        \midrule
        Online LSTM prediction
          &\includegraphics[width=\linewidth,height=\imgheight]{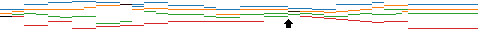}\\
        \midrule
        Offline BiLSTM prediction
          &\includegraphics[width=\linewidth,height=\imgheight]{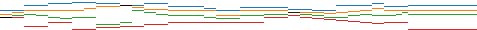}\\
        \multicolumn{2}{r}{\footnotesize (Audio available.\cref{fn:demo} Colors: \textcolor{tab-blue}{soprano}, \textcolor{tab-orange}{alto}, \textcolor{tab-green}{tenor}, \textcolor{tab-red}{bass}.)}
    \end{tabularx}
    \caption{Example of the Bach chorales dataset---\textit{Wer nur den lieben Gott läßt walten}, BWV 434, measures 1--5. The LSTM model makes two errors for the bass, as indicated by the arrow. The BiLSTM model gives a perfect prediction.}
    \label{fig:sample_bach}
\end{figure}

\section{Experiments}

\subsection{Implementation details}

We use a batch size of 16, a sequence length of 500 for training and a maximum sequence length of 2000 for validation and testing. We clip the time by 4096 time steps (i.e., roughly 170 quarter notes), the beat by 4096 beats, and durations by 192 time steps (i.e., 8 quarter notes). We randomly transpose the music by -5 to +6 semitones during training for data augmentation. We use the cross entropy loss with the Adam optimizer with $\alpha = 0.001$, $\beta_1 = 0.9$ and $\beta_2 = 0.999$~\cite{kingma2015adam}. We apply dropout~\cite{srivastava2014dropout} to prevent overfitting and layer normalization~\cite{ba2016layernorm} to speed up the training. All models are implemented in TensorFlow~\cite{abadi2016tensorflow} and experiments are run on NVIDIA GeForce RTX 2070s.

\subsection{Qualitative results and error analysis}

\begin{figure}
    \centering
    \small
    \def\imgheight{6.5ex}
    \begin{tabularx}{\linewidth}{m{.16\linewidth} X}
        Ground truth &\includegraphics[width=\linewidth,height=\imgheight]{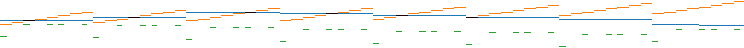}\\
        \midrule
        Online LSTM\hspace{1ex} prediction &\includegraphics[width=\linewidth,height=\imgheight]{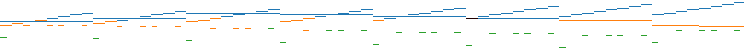}\\
        \midrule
        Offline BiLSTM prediction &\includegraphics[width=\linewidth,height=\imgheight]{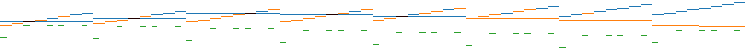}\\
        \multicolumn{2}{r}{\footnotesize (Audio available.\cref{fn:demo} Colors: \textcolor{tab-blue}{pulse wave I}, \textcolor{tab-orange}{pulse wave II}, \textcolor{tab-green}{triangle wave}.)}
    \end{tabularx}
    \caption{Hard excerpt in the game music dataset---\textit{Theme of Universe} from \textit{Miracle Ropit's Adventure in 2100}. Both models perform poorly when there is a sequence of short notes crossing a single long note.}
    \label{fig:sample_nes}
\end{figure}

\begin{figure}
    \centering
    \small
    \def\imgheight{8.5ex}
    \begin{tabularx}{\linewidth}{m{.16\linewidth} X}
        Ground truth
          &\includegraphics[width=\linewidth,height=\imgheight]{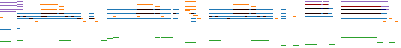}\\
        \midrule
        Online LSTM\hspace{1ex} prediction
          &\includegraphics[width=\linewidth,height=\imgheight]{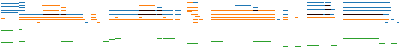}\\
        \midrule
        Offline BiLSTM prediction
          &\includegraphics[width=\linewidth,height=\imgheight]{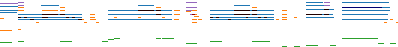}\\
        \multicolumn{2}{r}{\footnotesize (Audio available.\cref{fn:demo} Colors: \textcolor{tab-blue}{piano}, \textcolor{tab-orange}{guitar}, \textcolor{tab-green}{bass}, \textcolor{tab-red}{strings}, \textcolor{tab-purple}{brass}.)}
    \end{tabularx}
    \caption{Hard excerpt in the pop music dataset---\textit{Blame It On the Boogie} by The Jacksons. The BiLSTM model correctly identify and separate the overlapping guitar melody and piano chords, while the LSTM model fails in this case.}
    \label{fig:sample_lmd}
\end{figure}

We present in \cref{fig:sample_bach,fig:sample_musicnet,fig:sample_nes,fig:sample_lmd} several examples in the four datasets. Some representative cases include overlapping pitch ranges or chords for two polyphonic instruments (see~\cref{fig:sample_musicnet,fig:sample_lmd}), overlapping melodies and chords (see~\cref{fig:sample_lmd}) and a sequence of short notes crossing a single long note (see~\cref{fig:sample_nes}).

\subsection{Quantitative results}

We conduct an extensive empirical evaluation over different dataset and models in different settings. We present the results in \cref{tab:exp}.\footnote{Due to high computation cost, we report the oracle cases for the zone-based algorithm and MLP model on a subset of $100$ test samples, and omit the oracle case of the zone-based algorithm for the pop music dataset.} First, we notice the improved performance for the oracle cases on the MLP baseline. The large gap of performance is possibly because it predicts each note independently and the errors can propagate over time. This emphasizes the need to incorporate sequential models for this task. Moreover, the BiLSTM model outperforms its LSTM counterpart for most cases. This is reasonable as the BiLSTM model has access to the future information, which could, for example, help identify the direction of an arpeggio. Further, the LSTM and BiLSTM models outperform their Transformer counterparts---Transformer-Dec and Transformer-Enc, respectively---across all settings. However, the Transformer models benefits from faster inference speed at test time as compared to the LSTM models. Finally, we notice that the proposed models perform relatively poorly on the string quartets and game music datasets, possibly because the two violins in the string quartets dataset and the two pulse waves in the game music dataset are sometimes used interchangeably. We examine the use of pitch hints to help the models in the following section.

\begin{table}
    \centering
    \small
    \newcommand{\colsep}{\hspace{1.25em}}
    \begin{tabular}{ll@{\colsep}l@{\colsep}l@{\colsep}l}
        \toprule
        Model &Bach &String &Game &Pop\\
        \midrule
        \textbf{Online models}\\
        Zone-based                                   &73.14 &58.85 &43.67 &57.07\\
        MLP~\cite{devalk2019deepseparation}          &81.63 &29.85 &43.08$^*$ &33.50$^*$\\
        LSTM                                         &\textbf{93.02} &\textbf{67.43} &\textbf{50.22} &\textbf{74.14}\\
        Transformer-Dec                              &91.51 &57.03 &45.82 &62.14\\
        \cmidrule[.25pt](lr){1-5}
        Zone-based (oracle)                          &78.33 &66.89 &79.54$^*$ &$^\dag$\\
        MLP~\cite{devalk2019deepseparation} (oracle) &97.59 &58.16 &65.30 &44.62\\
        \midrule
        \textbf{Offline models}\\
        BiLSTM                                       &\textbf{97.13} &\textbf{74.3}8 &\textbf{52.93} &\textbf{77.23}\\
        Transformer-Enc                              &96.81 &58.86 &49.14 &66.57\\
        \midrule
        \multicolumn{5}{l}{\textbf{Online models (+entry hints)}}\\
        Closest-pitch                                &68.87 &50.69 &57.14 &47.45\\
        Closest-pitch (mono)                         &89.76 &42.82 &49.91 &32.28\\
        LSTM                                         &\textbf{92.70} &\textbf{62.64} &\textbf{62.11} &\textbf{74.19}\\
        Transformer-Dec                              &91.17 &62.12 &56.73 &67.19\\
        \midrule
        \multicolumn{5}{l}{\textbf{Offline models (+entry hints)}}\\
        BiLSTM                                       &\textbf{97.39} &\textbf{71.51} &\textbf{64.79} &\textbf{75.59}\\
        Transformer-Enc                              &93.81 &56.72 &54.67 &67.23\\
        \bottomrule
    \end{tabular}\\
    \raggedright{\footnotesize $^*$Reported on a subset of $100$ test samples due to high computation cost.\\$^\dag$Omitted due to high computation cost.}
    \caption{Comparison of our proposed models and baseline algorithms. Performance is measured in accuracy (\%).}
    \label{tab:exp}
\end{table}

\begin{table}
    \centering
    \small
    \newcommand{\colsep}{\hspace{1em}}
    \begin{tabular}{l@{\colsep}l@{\colsep}l@{\colsep}ll@{\colsep}l@{\colsep}l@{\colsep}l}
        \toprule
        Emb        &Dur        &EH          &PH          &Bach &String &Game &Pop\\
        \midrule
                   &           &            &            &92.10 &37.29 &43.89 &58.78\\
        \checkmark &           &            &            &93.02 &67.43 &50.22 &74.14\\
        \checkmark &\checkmark &            &            &\textbf{96.17} &66.96 &51.38 &\textbf{78.17}\\
        \checkmark &           &\checkmark  &            &92.70 &62.64 &62.11 &74.19\\
        \checkmark &\checkmark &\checkmark  &            &95.95 &68.17 &63.35 &74.74\\
        \checkmark &           &            &\checkmark  &92.87 &\textbf{70.20} &\textbf{67.45} &75.89\\
        \bottomrule
    \end{tabular}
    \caption{Effects of input features to the online LSTM model. Performance is measured in accuracy (\%). Abbreviations: `Emb'---pitch, beat and position embedding, `Dur'---duration, `EH'---entry hints, `PH'---pitch hints.}
    \label{tab:exp_features}
\end{table}

\begin{figure*}
    \centering
    \def\imgheight{6ex}
    \small
    \begin{tabularx}{\linewidth}{m{.01\linewidth} X}
        (a) &\includegraphics[width=\linewidth,height=\imgheight]{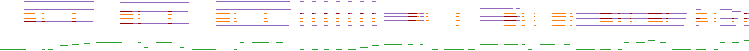}\\
        \midrule
        (b) &\includegraphics[width=\linewidth,height=\imgheight]{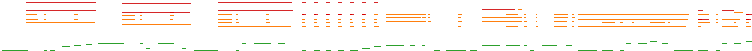}\\
        \midrule
        (c) &\includegraphics[width=\linewidth,height=\imgheight]{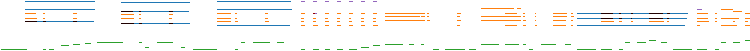}\\
        \multicolumn{2}{r}{\footnotesize (Audio available.\cref{fn:demo} Colors: \textcolor{tab-blue}{piano}, \textcolor{tab-orange}{guitar}, \textcolor{tab-green}{bass}, \textcolor{tab-red}{strings}, \textcolor{tab-purple}{brass}.)}
    \end{tabularx}
    \caption{\textit{Quando Quando Quando} by Tony Renis---(a) original instrumentation and the versions produced by (b) the online LSTM model without entry hints and (b) the offline BiLSTM model with entry hints. The LSTM model assigns the chords to the guitar, the most common instrument in the pop music dataset except the high pitches, which are assigned to the strings. The BiLSTM model is able to separate the long chords from the short ones and assigns the former to the piano.}
    \label{fig:sample_instrumentation}
\end{figure*}

\subsection{Effect of input features}

In order to compare the effectiveness of different input features, we also report in \cref{tab:exp_features} the performance for the LSTM model with different input features. First of all, pitch, note and beat embedding leads to improvements on all datasets, especially significant on the string quartet (30\% gain) and pop music (15\% gain) datasets. Second, entry hints improve the performance by 10\% for the game music dataset, which is possibly because it helps disambiguate the two interchangeable pulse wave tracks. Interestingly, they have negative impacts on the Bach chorales and pop music datasets. Third, duration inputs are always helpful and help achieve the highest accuracy on the Bach chorales and pop music datasets. For example, durations would be critical in distinguishing the overlapping guitar melody and piano chords in the example shown in \cref{fig:sample_lmd}. Last, pitch hints improve the performance for all datasets but Bach chorales, possibly because the vocal ranges for SATB are strict in chorales. Pitch hints help achieve the highest accuracies for the string quartets and game music datasets as they help disambiguate interchangeable tracks.

\subsection{Effects of time encoding}

In this experiment, we examine the effects of time encoding. In particular, we consider four variants---(1) raw time as a number, (2) raw beat and position as two numbers, (3) time embedding and (4) beat and position embedding (see \cref{sec:model} for the definition of beat and position). We report in \cref{tab:ablation} the results and we can see that using raw time gives the worst performance. Interestingly, the other three encoding strategies achieve comparable performance.

\begin{table}
    \centering
    \small
    \begin{tabular}{lllll}
        \toprule
        Strategy &Bach &String &Game &Pop\\
        \midrule
        \textbf{Time encoding}\\
        Raw time                    &91.97 &37.26 &44.10 &37.92\\
        Raw beat and position       &\textbf{93.13} &66.72 &48.60 &68.42\\
        Time embedding              &92.21 &\textbf{68.31} &49.32 &70.64\\ 
        Beat and position emb.      &93.02 &67.43 &\textbf{50.22} &\textbf{74.14}\\
        \midrule
        \textbf{Data augmentation}\\
        No augmentation             &\textbf{93.03} &\textbf{69.36} &49.03 &70.73\\
        Light augmentation          &92.85 &68.66 &46.38 &71.10\\
        Strong augmentation         &93.02 &67.43 &\textbf{50.22} &\textbf{74.14}\\
        \bottomrule
    \end{tabular}
    \caption{Comparisons of time encoding and data augmentation strategies for the online LSTM model. Performance is measured in accuracy (\%).}
    \label{tab:ablation}
\end{table}

\subsection{Effects of data augmentation}

In this experiment, we compare the following three strategies of data augmentation---(1) no augmentation, (2) \emph{light augmentation}, where each song is randomly transposed by -1 to +1 semitone during training and (3) \emph{strong augmentation}, where each song is randomly transposed by -5 to +6 semitones during training. We report in \cref{tab:ablation} the results. We can see that data augmentation is generally harmful for the Bach chorales and string quartets datasets, possibly because classical music has strict rules on the pitch ranges of voices and instruments. However, for game and pop music datasets, where rules on keys and pitch ranges in classical music are loosened, the models yield better performance with proper data augmentation.

\section{Discussion}

In \cref{fig:sample_instrumentation}, we depict the original instrumentation of the song \textit{Quando Quando Quando} alongside the instrumentations generated by our best performing models for both the online and offline settings. While neither model produces an instrumentation identical to that of the original, both produce instrumentations that ``cluster'' notes similarly to the original and are reasonable rearrangements of the song. This indicates a fundamental ambiguity of the task, though we note that such ambiguity is less present in some genres than others---our models are able to achieve high accuracy on the Bach chorales dataset despite its small size. However, for larger and more diverse datasets (e.g., the pop music dataset), accuracy might not be the best metric for measuring the performance of the models, and we plan to include human evaluations in future work.

One limitation of this work lies in the generalizability to real keyboard music since the downmixed music might not be playable on a keyboard, e.g., having more than ten concurrent notes or impossible fingering. Moreover, we did not use the MIDI velocity information in our models, and it could provide an additional signal for separation.

Finally, in addition to its immediate musical applications, we believe that our proposed part separation task may be useful for large-scale pre-training of symbolic music models. Pre-training music generation models on large, heterogeneous music corpora has already been observed to improve performance~\cite{donahue2019lakhnes,hung2019improving}. Given that our proposed task represents an additional source of musical knowledge supervision, we speculate that additionally pre-training on this task could improve performance for many downstream tasks, e.g., genre classification and melody extraction.

\section{Conclusion}

In this paper, we have proposed a new task of part separation in multitrack music and examined its feasibility under both the online and offline settings. Through a comprehensive empirical evaluation over four diverse datasets, we showed the effectiveness of our proposed models against various baselines. We also presented promising results for applying part separation models to automatic instrumentation. Moreover, we discussed the fundamental ambiguity and limitations of the task and future research directions.

\bibliography{ref}

\end{document}